\documentclass[12pt]{article}
\usepackage{amsmath,amsfonts,amssymb}  
\makeatletter
\def\numberbysection{\@addtoreset{equation}{section}
    \def\theequation{\thesection.\arabic{equation}}}
\makeatother

\numberbysection

\newcommand{\be}{\begin{eqnarray}}
\newcommand{\ee}{\end{eqnarray}}
\newcommand{\non}{\nonumber}
\newcommand{\id}{\mathbb{I}}

\newcommand{\str}{\mathop{\rm str}\nolimits}

\newcommand{\diag}{\mathop{\rm diag}\nolimits}

\newcommand{\nonu}{\nonumber \\}
\newcommand{\mb}[1]{\hs{4}\mbox{#1}\hs{4}}
\newcommand{\hs}[1]{\hspace{#1 mm}}

\def\cT{{\cal T}}  
\newcommand{\II}{{\mathbb I}}
\newcommand{\wh}[1]{\widehat{#1}}
\newcommand{\atopn}[2]{\genfrac{}{}{0pt}{}{#1}{#2}}

\begin{document}

\begin{titlepage}
\strut\hfill [arXiv:0810.5015]\\
\strut\hfill UMTG--261\\
\strut\hfill LAPTH-1285/08\\
\vspace{.5in}
\begin{center}

\LARGE Analytical Bethe ansatz for the open\\
       AdS/CFT $SU(1|1)$ spin chain\\[1.0in]
\large Rafael I. Nepomechie\footnote{
       Physics Department, P.O. Box 248046, University of Miami,
       Coral Gables, FL 33124 USA;  nepomechie@physics.miami.edu}      
      and Eric Ragoucy\footnote{
       Laboratoire d'Annecy-le-Vieux de Physique Th\'eorique
       LAPTH, UMR 5108, CNRS and Universit\'e de Savoie, B.P. 110, F-74941 Annecy-le-Vieux 
       Cedex, France;  ragoucy@lapp.in2p3.fr}\\

\end{center}

\vspace{.5in}

\begin{abstract}
We prove an inversion identity for the open AdS/CFT $SU(1|1)$ quantum
spin chain which is exact for finite size.  We use this identity,
together with an analytic ansatz, to determine the eigenvalues of the
transfer matrix and the corresponding Bethe ansatz equations. We also 
solve the closed chain by algebraic Bethe ansatz.
\end{abstract}

\end{titlepage}

\setcounter{footnote}{0}

\section{Introduction}\label{sec:intro}

The factorized $SU(2|2)$-invariant bulk $S$-matrix \cite{St, Be1, AFZ}
plays a central role in understanding integrability in the closed
string/spin chain sector of AdS/CFT. Indeed, this $S$-matrix can be
used to derive \cite{Be1, MM, dL} the all-loop asymptotic Bethe ansatz
equations (BAEs) \cite{BS} and to compute finite-size effects
\cite{BJ}.  The corresponding factorized boundary $S$-matrices
\cite{HM}-\cite{CY} should play a parallel role in the open
string/spin chain sector. \footnote{For earlier work, see 
\cite{CWW}-\cite{BL} and references therein.}
The commuting open-chain transfer matrix,
which is constructed from both bulk and boundary $S$-matrices, was
recently formulated (following \cite{Sk}) in \cite{MN}.  Determining
the eigenvalues of this transfer matrix, which has yet to be
accomplished, is the key technical step to determining the
corresponding all-loop BAEs.

A simpler $SU(1|1)$-invariant bulk $S$-matrix (which is in fact a
submatrix of the $SU(2|2)$ $S$-matrix) was found in \cite{BS, Be2},
and a corresponding boundary $S$-matrix was found in \cite{MN}.  The
purpose of this paper is to determine the eigenvalues and BAEs of the
open-chain transfer matrix constructed from these $S$-matrices
\cite{MN}.  We expect that this computation will serve as a useful
warm-up exercise for the more realistic $SU(2|2)$ case.

An essential element of our computation is the derivation of an 
inversion identity for the transfer matrix $t(p)$, namely,
\be
t(p)\, t(-p) = \Lambda_{0}(p)\, \Lambda_{0}(-p)\,  \id \,, 
\label{inversionIntro}
\ee
where $\Lambda_{0}(p)$ is a known scalar function, which is an exact
equation for a chain of finite size $L$.  Similar relations (although not
necessarily exact for finite size) have long been known for various
lattice models \cite{Str, Ba}.  Together with a suitable analytic
ansatz (see, e.g., \cite{Re}-\cite{RS}), the transfer matrix eigenvalues
and associated BAEs can then be obtained.

The outline of this paper is as follows.  In Section
\ref{sec:transfer} we review the construction of the transfer matrix.
In Section \ref{sec:BA} we obtain the eigenvalues and BAEs of the
transfer matrix.  We also work out the weak-coupling $( g \rightarrow
0)$ limit.  We briefly discuss these results in Section
\ref{sec:discuss}.  There are several appendices.  In Appendix
\ref{app:closed} we solve the closed chain (using algebraic Bethe
ansatz), since we use the form of this solution to help formulate the
analytical ansatz for the open chain.  We compute the pseudovacuum
eigenvalue in Appendix \ref{app:pseudovacuum}.  We present a proof of
the inversion identity in Appendix \ref{app:inversion}, and we prove 
in Appendix \ref{app:cross}  a crossing-like identity for the transfer 
matrix, the dressing function being compatible with this identity.

\section{The transfer matrix and its properties}\label{sec:transfer}

The basic building blocks from which the open-chain transfer matrix is
constructed are bulk and boundary $S$-matrices.  We begin this Section
by reviewing these $S$-matrices.  We then briefly review the
construction of the transfer matrix and present its important
properties.

\subsection{Bulk $S$-matrix}

The $SU(1|1)$ bulk $S$-matrix is given by \cite{BS, Be2} \footnote{We 
are not concerned here with overall scalar factors.}
\be
S(p_{1}, p_{2}) = \left( \begin{array}{cccc}
x^{+}_{1}-x^{-}_{2} &0            &0           &0  \\
0                 &x^{-}_{1}-x^{-}_{2}    
&(x^{+}_{1}-x^{-}_{1})\frac{\omega_{2}}{\omega_{1}}  &0  \\
0                 
&(x^{+}_{2}-x^{-}_{2})\frac{\omega_{1}}{\omega_{2}}  &x^{+}_{1}-x^{+}_{2}  &0 \\
0                 &0            &0           &x^{-}_{1}-x^{+}_{2}
\end{array} \right) \,,
\label{bulkS}
\ee
where $x_{i}^{\pm} = x^{\pm}(p_{i})$, $\omega_{i} = \omega(p_{i})$. 
This $S$-matrix is regular,
\be
S(p, p) \propto {\cal P} \,,
\ee
where ${\cal P}$ is the graded permutation matrix,
\be
{\cal P} = \sum_{i,j=1}^{2} (-1)^{p(i) p(j)} e_{i\, j} \otimes 
e_{j\, i} \,,
\ee
where $e_{i j}$ is the usual elementary $2 \times 2$ matrix whose
$(i,j)$ matrix element is 1, and all others are zero; and the parity
assignments are $p(1)=0$, $p(2)=1$. 
It has the unitarity property
\be
S_{12}(p_{1},p_{2})\,S_{21}(p_{2},p_{1})\ =\ 
(x^+_{1}-x^-_{2})(x^+_{2}-x^-_{1})\,\II\otimes\II 
\label{eq:Sunit}
\ee
where $S_{21} = {\cal P}_{12}\, S_{12}\, {\cal P}_{12}$
and $\id$ is the two-dimensional identity matrix; and it satisfies
the graded Yang-Baxter equation (YBE) \cite{KS}
\be
S_{12}(p_{1}, p_{2})\, S_{13}(p_{1}, p_{3})\, S_{23}(p_{2}, p_{3})\ =
S_{23}(p_{2}, p_{3})\, S_{13}(p_{1}, p_{3})\, S_{12}(p_{1}, p_{2}) \,,
\label{YBE}
\ee
where 
\be
S_{12} = S \otimes \id\,, \quad 
S_{13} = {\cal P}_{23}\, S_{12}\, {\cal P}_{23}\,, \quad
S_{23} = {\cal P}_{12}\, S_{13}\, {\cal P}_{12}\,,
\ee
and ${\cal P}_{12} = {\cal P} \otimes \id$, ${\cal P}_{23} = \id \otimes {\cal P}$. 
The bulk $S$-matrix does not have, to our knowledge, true crossing 
symmetry. \footnote{For the $SU(M|N)$ Yangian $S$-matrix $S(u) = u 
\id + i {\cal P}$, one can argue that crossing symmetry 
$$S^{t_{1}}_{12}(u + \eta) = A\, S_{12}(u)\, A^{-1}$$
is possible only for $|M-N|=2$. Indeed, canceling the $u$'s and squaring both sides, one obtains
$$\left(\eta \id + i{\cal P}^{t_{1}} \right)^{2} = \left(i A\, {\cal 
P}\,  A^{-1}\right)^{2} = -\id \,. $$
Using the fact $\left(  {\cal P}^{t_{1}} \right)^{2} = (M-N) {\cal 
P}^{t_{1}}$, it follows that $\eta = \pm i\,, M-N = \pm 2$.}
However, it does obey the ``crossing-like'' relations
\begin{eqnarray}
&& S^{t_{1}}_{12}(p_{1},p_{2})\ =\ \sigma^y_{1}\,S_{12}(p_{1},p_{2})
\,\sigma^y_{1}\Big|_{x^+_{2}\leftrightarrow x^-_{2}} \,,
\label{eq:Scross1}\\
&& S^{t_{2}}_{12}(p_{1},p_{2})\ =\ \sigma^y_{2}\,S_{12}(p_{1},p_{2})
\,\sigma^y_{2}\Big|_{x^+_{1}\leftrightarrow x^-_{1}} \,,
\label{eq:Scross2}
\end{eqnarray}
which imply PT symmetry
\begin{eqnarray}
S^{t_{1}t_{2}}_{12}(p_{1},p_{2})\ =\ 
-\sigma^y_{1}\,\sigma^y_{2}\,S_{21}(p_{2},p_{1})
\,\sigma^y_{2}\,\sigma^y_{1}\,, 
\label{eq:Scross12}
\end{eqnarray}
where $t_{i}$ denotes super-transposition \cite{Ritt} in the $i^{th}$ space,
\begin{equation}
A^t = \sum_{i,j=1}^{2} (-1)^{p(j)\big(p(i)+p(j)\big)}\,A_{ij}\,e_{ji}
\mb{if} 
A = \sum_{i,j=1}^{2} A_{ij}\,e_{ij}\,,
\end{equation}
 and
the subscripts $x^+_{\ell}\leftrightarrow x^-_{\ell}$, 
indicate that one has to exchange $x^+(p_{\ell})$ with $x^-(p_{\ell})$
in the $S$-matrix. Moreover, $\sigma^y_{i}$ is the second Pauli matrix in space $i$.

As noted by Beisert and Staudacher \cite{BS}, the YBE 
is satisfied without imposing any constraint between $x^{+}(p)$ and 
$x^{-}(p)$, and without specifying $\omega(p)$. For convenience, we 
henceforth set 
\be
\omega(p) = 1\,,
\label{omega}
\ee
or equivalently, we gauge $\omega(p)$ away by performing the gauge 
transformation \footnote{This transformation does not modify the eigenvalues
of the transfer matrix, which we define in Sec. \ref{subsec:transfer}.}
\begin{equation}
S_{12}(p_{1},p_{2})\ \to\ 
G_{1}(p_{1})\,G_{2}(p_{2})\,S_{12}(p_{1},p_{2})\,G_{2}(p_{2})^{-1}\, G_{1}(p_{1})^{-1}
\mb{with} G(p)=\left(\begin{array}{cc} \omega(p) & 0 \\ 0 
&1\end{array}\right) \,.
\end{equation}

If we regard $x^{\pm}(p)$ as the usual functions satisfying
\be
x^{+}+\frac{1}{x^{+}}-x^{-}-\frac{1}{x^{-}} = \frac{i}{g}\,, \qquad 
\frac{x^{+}}{x^{-}} = e^{i p} \,,
\ee 
then the weak-coupling limit corresponds to setting
\be
x^{\pm} = \frac{1}{g}(u \pm i/2) \,,
\label{WeakCoupling}
\ee
and then letting $g \rightarrow 0$.  In this limit, the $S$-matrix
(\ref{bulkS}) evidently reduces \cite{BS} to the well-known $SU(1|1)$
``Yangian'' $S$-matrix,
\be
S(p_{1}, p_{2}) \rightarrow \frac{1}{g} S(u_{1}-u_{2})\,, \qquad
S(u_{1}-u_{2}) \equiv 
(u_{1} - u_{2}) \id + i {\cal P} \,,
\label{bulkSweak}
\ee
where $u_{i} = u(p_{i})$.

\subsection{Boundary $S$-matrices}

The right boundary $S$-matrix is given by the diagonal $2 \times 2$ matrix \cite{MN}
\be
R^{-}(p) = \diag\left(a - x^{+}(p)\,, a + x^{-}(p) \right) \,,
\label{Rminus}
\ee
where $a$ is an arbitrary boundary parameter. It satisfies the 
(right) boundary Yang-Baxter equation \cite{Ch, GZ}
\be
S_{12}(p_{1}, p_{2})\, R_{1}^{-}(p_{1})\, S_{21}(p_{2}, -p_{1})\, 
R_{2}^{-}(p_{2}) = 
R_{2}^{-}(p_{2})\, S_{12}(p_{1}, -p_{2})\, R_{1}^{-}(p_{1})\, 
S_{21}(-p_{2}, -p_{1}) 
\label{BYBE}
\ee 
without imposing any constraint between $x^{+}(p)$ and $x^{-}(p)$ 
other than \cite{HM} 
\be
x^{\pm}(-p) = - x^{\mp}(p) \,.
\label{negation}
\ee 
Moreover, the left boundary $S$-matrix is given by \cite{MN}
\be
R^{+}(p) =  R^{-}(-p)\Big\vert_{a \mapsto b} = 
\diag\left(b + x^{-}(p)\,, b - x^{+}(p) \right) \,,
\label{Rplus}
\ee
where $b$ is another arbitrary boundary parameter.

In the weak-coupling limit (\ref{WeakCoupling}), the boundary 
$S$-matrices reduce to
\be
R^{-}(p) &\rightarrow&  \frac{1}{g} R^{-}(u)\,, \qquad
R^{-}(u)  \equiv (\alpha - i/2) \id - u\, \sigma^{z}
\,, \non \\
R^{+}(p) &\rightarrow&  \frac{1}{g} R^{+}(u)\,, \qquad
R^{+}(u) \equiv (\beta - i/2) \id + u\, \sigma^{z}  
\,,
\label{Rweak}
\ee
where we have set $a=\alpha/g$ and $b=\beta/g$.

\subsection{Transfer matrix}\label{subsec:transfer}

The open-chain transfer matrix is constructed from the bulk and 
boundary $S$-matrices as follows \cite{Sk, MN}
\be
t(p\,; \{p_{\ell}\}) &=& \str_{0} R_{0}^{+}(p)\, {\cal T}^{-}_{0}(p\,; 
\{p_{\ell}\})  \non \\
&=& \str_{0} R_{0}^{+}(p)\, T_{0}(p\,; \{p_{\ell}\})\,
R_{0}^{-}(p)\,  \widehat T_{0}(p\,; \{p_{\ell}\}) \,,
\label{transfer}
\ee 
where $T$ and $\widehat T$ are a pair of monodromy matrices
\be
T_{0}(p\,; \{p_{\ell}\}) &=& S_{0 L}(p,p_{L}) \cdots S_{0 1}(p,p_{1}) \,,
\non \\
\widehat T_{0}(p\,; \{p_{\ell}\}) &=& S_{1 0}(p_{1}, -p) \cdots 
S_{L 0}(p_{L}, -p) \,,
\label{monodromy}
\ee
$\{p_{1}, \ldots, p_{L}\}$ are arbitrary ``inhomogeneities''
associated with each of the $L$ quantum spaces, the auxiliary
space is denoted here by $0$, and $\str$ denotes 
supertrace \cite{FrKl,NaSch,Kac77}:
\begin{equation}
str(A)=\sum_{i=1}^{2}(-1)^{p(i)} A_{ii} = 
A_{11}-A_{22}\mb{for} A=\sum_{i,j=1}^2 A_{ij}\,e_{ij}\,.
\end{equation}

The transfer matrix is constructed to have the commutativity
property
\be
\left[ t(p\,; \{p_{\ell}\}) \,, t(q\,; \{p_{\ell}\}) \right] = 0 
\label{commutativity}
\ee
for arbitrary values of $p$ and $q$. 

The transfer matrix also obeys the exact inversion identity
\be
t(p\,; \{p_{\ell}\})\, t(-p\,; \{p_{\ell}\}) 
= \Lambda_{0}(p\,; \{p_{\ell}\})\, \Lambda_{0}(-p\,; \{p_{\ell}\})\, \id \,, 
\label{inversion}
\ee
where $\Lambda_{0}(p\,; \{p_{\ell}\})$ is the pseudovacuum eigenvalue.
This identity can easily be verified numerically for small values of 
$L$, and we prove it by recursion on $L$ in Appendix \ref{app:inversion}.

In the weak-coupling limit  (\ref{WeakCoupling}) with also
$x^{\pm}_{\ell} = \frac{1}{g}(u_{\ell} \pm i/2)$, the transfer matrix 
(\ref{transfer}) becomes \footnote{We suppress the overall factor 
$1/g^{2L+2}$.}
\be
t(u\,; \{u_{\ell}\}) = \str_{0} R_{0}^{+}(u) 
S_{0 L}(u - u_{L}) \cdots S_{0 1}(u - u_{1})
R_{0}^{-}(u) 
S_{1 0}(u + u_{1}) \cdots 
S_{L 0}(u + u_{L}) \,,\ 
\label{transferWeak}
\ee
where the bulk and boundary $S$-matrices are now given by 
(\ref{bulkSweak}), (\ref{Rweak}), respectively.

\section{Analytical Bethe ansatz}\label{sec:BA}

As shown in Appendix \ref{app:pseudovacuum}, the pseudovacuum state
consisting of all spins up
\be
\Omega=\underbrace{e_{1}\otimes e_{1}\otimes\ldots\otimes e_{1}}_{L}
\mb{where} e_{1}= \left( {1 \atop 0} \right)
\label{pseudovacuum}
\ee
is an eigenstate of the open-chain transfer matrix (\ref{transfer}), with 
eigenvalue
\be
\Lambda_{0}(p\,; \{p_{\ell}\}) &=& \frac{x^+(p) + x^-(p)}{2x^+(p)}\,\left\{
(a-x^+(p))(b+x^+(p))\prod_{\ell=1}^L 
(x^+(p)-x^-(p_{\ell}))(x^+(p)+x^+(p_{\ell}))
\right. \non \\
&&\quad
 \left.-(a+x^+(p))(b-x^+(p))\prod_{\ell=1}^L 
 (x^+(p)-x^+(p_{\ell}))(x^+(p) + x^-(p_{\ell}))
\right\} \,.
\label{Lambda0}
\ee

We make the ``analytical ansatz'' \cite{Re}-\cite{RS} that every eigenvalue
of the transfer matrix can be expressed as an appropriately
``dressed'' pseudovacuum eigenvalue,
\be
\Lambda(p\,; \{p_{\ell}\},\{\lambda_{j}\}) = \Lambda_{0}(p\,; \{p_{\ell}\})\,
A(p\,; \{\lambda_{j}\}) \,.
\label{Lambda}
\ee
In order to determine the ``dressing factor'' $A(p\,; 
\{\lambda_{j}\})$, we make use of the inversion identity 
(\ref{inversion}), which implies a corresponding identity for the 
eigenvalues,
\be
\Lambda(p\,; \{p_{\ell}\},\{\lambda_{j}\})\, \Lambda(-p\,; 
\{p_{\ell}\},\{\lambda_{j}\}) 
= \Lambda_{0}(p\,; \{p_{\ell}\})\, \Lambda_{0}(-p\,; \{p_{\ell}\}) \,.
\ee
It follows that the dressing factor must satisfy the constraint
\be
A(p\,; \{\lambda_{j}\})\, A(-p\,; \{\lambda_{j}\}) = 1 \,.
\label{dressingConstraint}
\ee
A natural conjecture is that the open-chain dressing factor 
$A(p\,; \{\lambda_{j}\})$ can be 
expressed in terms of the closed-chain dressing factor 
$A^{(c)}(p\,; \{\lambda_{j}\})$ (\ref{dressingClosed}),
\be
A(p\,; \{\lambda_{j}\})  = \frac{A^{(c)}(p\,; 
\{\lambda_{j}\})}{A^{(c)}(-p\,; \{\lambda_{j}\})} 
=  \prod_{j=1}^{M} \left( \frac{x^{-}(p)-x^{+}(\lambda_{j})}
{x^{+}(p)-x^{+}(\lambda_{j})} \right)
\left( \frac{x^{-}(p)+x^{+}(\lambda_{j})}
{x^{+}(p)+x^{+}(\lambda_{j})} \right) \,,
\label{dressing}
\ee
which evidently is consistent with the constraint
(\ref{dressingConstraint}).  
As discussed in Appendix \ref{app:cross}, the dressing factor
(\ref{dressing}) also obeys the crossing-like relation
(\ref{crossinglikereltn2}), which provides a further consistency
check.
The dressing factor (\ref{dressing}) obviously has poles
at $p=\lambda_{j}$.  The requirement that the eigenvalues
(\ref{Lambda}) be analytic implies that $\lambda_{j}$ must satisfy the
open-chain BAEs
\be
& & \left(\frac{a-x^+(\lambda_{j})}{a+x^+(\lambda_{j})}\right)
\left(\frac{b+x^+(\lambda_{j})}{b-x^+(\lambda_{j})}\right)
\prod_{\ell=1}^{L}\left( \frac{x^{+}(\lambda_{j})-x^{-}(p_{\ell})}
{x^{+}(\lambda_{j})-x^{+}(p_{\ell})} \right)
\left( \frac{x^{+}(\lambda_{j})+x^{+}(p_{\ell})}
{x^{+}(\lambda_{j})+x^{-}(p_{\ell})} \right) = 1\,, \non \\
& & \qquad j = 1, \ldots, M\,, \quad M = 0, 1, \ldots, L\,.
\label{BAE}
\ee
We have checked the completeness of this solution numerically for up
to $L=4$.

In the weak-coupling limit (\ref{WeakCoupling}), corresponding to the
transfer matrix (\ref{transferWeak}),
the solution (\ref{Lambda0}), (\ref{Lambda}), (\ref{dressing}), 
(\ref{BAE}) becomes
\be
\Lambda(u\,; \{u_{\ell}\}, \{\mu_{j}\})
= \Lambda_{0}(u\,; \{u_{\ell}\})\, 
\prod_{j=1}^{M} \left( \frac{u -\mu_{j} -i}{u -\mu_{j}} \right)
\left( \frac{u +\mu_{j}}{u +\mu_{j}+i} \right) \,, 
\ee
where
\be
\Lambda_{0}(u\,; \{u_{\ell}\}) &=& \frac{u}{2(2u+i)}\Big\{
(2\alpha -2u-i)(2\beta+2u+i)\prod_{\ell =1}^{L}\left( u - u_{\ell} + i \right)
\left( u + u_{\ell} + i \right) \non\\
& & - (2\alpha +2u+i)(2\beta-2u-i)\prod_{\ell=1}^{L}\left(u - u_{\ell} 
\right) \left(u + u_{\ell} \right) \Big\} \,,
\ee 
and the BAEs are given by
\be
\left( \frac{2\alpha -2\mu_{j}-i}{2\alpha +2\mu_{j}+i} \right)
\left( \frac{2\beta +2\mu_{j}+i}{2\beta -2\mu_{j}-i} \right)
\prod_{\ell=1}^{L}\left( \frac{\mu_{j}-u_{\ell}+i}
{\mu_{j}-u_{\ell}} \right)
\left( \frac{\mu_{j}+u_{\ell}+i}
{\mu_{j}+u_{\ell}} \right)= 1\,,
\ee 
where we have also set $\lambda_{j} =\mu_{j}/g$.

\section{Discussion}\label{sec:discuss}

We have found an exact inversion identity for the open-chain transfer
matrix constructed from the $SU(1|1)$ $S$-matrix \cite{BS, Be2} and
the corresponding boundary $S$-matrices \cite{MN}.  We have used this
inversion identity to help determine the transfer matrix eigenvalues
(\ref{Lambda0}), (\ref{Lambda}), (\ref{dressing}) and the associated
BAEs (\ref{BAE}).  These BAEs are evidently of the free-Fermion type,
as the various Bethe roots are not coupled.

While it should also be possible to obtain these results via algebraic
Bethe ansatz, we have instead pursued here the analytical Bethe ansatz
approach, since the latter should be more manageable for the $SU(2|2)$
case.  An inversion identity may also hold for that case.  Indeed, we have
verified this numerically in the weak-coupling limit with $R^{\pm}(u)
= \id$.

Our proof of the inversion identity (for the $SU(1|1)$ case) relies on
recursion on the size of the chain.  It would be interesting to find a
more direct proof.  Unfortunately, conventional
fusion techniques for graded $SU(n|m)$ chains 
\cite{Tsuboi,selene} do not seem to work for
the $n=m$ case (see e.g., \cite{Na}) which we consider here.

\section*{Acknowledgments}
This work was initiated when one of the authors (E.R.) visited the
University of Miami: he wishes to thank the University and the members
of the Physics Department for warm hospitality and partial support
during his stay.  The other author (R.N.) thanks the Galileo Galilei
Institute for Theoretical Physics for hospitality and the INFN for
partial support during the completion of this work.  This work was
supported in part by the National Science Foundation under Grants
PHY-0244261 and PHY-0554821.

\begin{appendix}

\section{The closed $SU(1|1)$ spin chain\label{app:closed}}

The commuting transfer matrix for the closed $SU(1|1)$ spin chain is
given by
\be
t^{(c)}(p\,; \{p_{\ell}\}) = \str_{0} T_{0}(p\,; \{p_{\ell}\}) \,,
\label{transferClosed}
\ee
where $T$ is the monodromy matrix defined in (\ref{monodromy}), which
satisfies the fundamental relation
\be
S_{0 0'}(p, q)\, T_{0}(p\,; \{p_{\ell}\})\,
T_{0'}(q\,; \{p_{\ell}\})  = T_{0'}(q\,; \{p_{\ell}\})\,
T_{0}(p\,; \{p_{\ell}\})\, S_{0 0'}(p, q) \,.
\label{FR}
\ee

The eigenvalues of the transfer matrix can easily be determined by 
algebraic Bethe ansatz. As usual \cite{Fa}, we write the monodromy 
matrix as a matrix in the auxiliary space
\be
T_{0}(p\,; \{p_{\ell}\}) = \left( \begin{array}{cc}
A(p\,; \{p_{\ell}\}) & B(p\,; \{p_{\ell}\}) \\
C(p\,; \{p_{\ell}\}) & D(p\,; \{p_{\ell}\})
\end{array} \right) \,.
\ee
The pseudovacuum state (\ref{pseudovacuum}) consisting of all spins up
is an eigenstate of both $A(p\,; \{p_{\ell}\})$ and
$D(p\,; \{p_{\ell}\})$,
\be
A(p\,; \{p_{\ell}\})\, \Omega &=& \prod_{\ell=1}^{L}\left( x^{+}(p)-x^{-}(p_{\ell}) 
\right) \Omega \,, \non \\
D(p\,; \{p_{\ell}\})\, \Omega &=& \prod_{\ell=1}^{L}\left( x^{+}(p)-x^{+}(p_{\ell}) 
\right) \Omega \,.
\ee 

The fundamental relation (\ref{FR}) implies the relations 
\footnote{We remark that the $B$ operators do not commute:
\be
B(p\,; \{p_{\ell}\}) B(q\,; \{p_{\ell}\}) = \left(
\frac{x^{-}(p)-x^{+}(q)}{x^{-}(q)-x^{+}(p)} \right) 
B(q\,; \{p_{\ell}\}) B(p\,; \{p_{\ell}\}) \,. \non
\ee
A similar non-commutativity has been observed in other graded models, 
see e.g. \cite{Kulish,BR}.}
\be
A(p\,; \{p_{\ell}\}) B(q\,; \{p_{\ell}\}) &=& f(p, q) B(q\,; \{p_{\ell}\})  A(p\,; \{p_{\ell}\})
+ g(p, q) B(p\,; \{p_{\ell}\})  A(q\,; \{p_{\ell}\}) 
\qquad\label{AB}\\
D(p\,; \{p_{\ell}\}) B(q\,; \{p_{\ell}\}) &=& f(p, q) B(q\,; 
\{p_{\ell}\})  D(p\,; \{p_{\ell}\})
+ g(p, q) B(p\,; \{p_{\ell}\})  D(q\,; \{p_{\ell}\})  
\qquad\label{DB}
\ee
where
\be
f(p,q) = \frac{x^{-}(p)-x^{+}(q)}{x^{+}(p)-x^{+}(q)} \,, \qquad
g(p,q) = \frac{x^{+}(q)-x^{-}(q)}{x^{+}(p)-x^{+}(q)} \,.
\ee 
Note that the {\it same} functions $f(p,q)$, $g(p,q)$ appear in both 
(\ref{AB}) and (\ref{DB}).

Consider the state obtained by applying a set of $B$ operators 
with arguments $\lambda_{1}, \ldots, \lambda_{M}$ on the 
pseudovacuum state,
\be
| \lambda_{1}, \ldots, \lambda_{M}\rangle = 
B(\lambda_{1}\,; \{p_{\ell}\}) \ldots B(\lambda_{M}\,; \{p_{\ell}\})\, \Omega 
\,.
\ee
This state is an eigenstate of the transfer matrix 
$t^{(c)}(p\,; \{p_{\ell}\}) = A(p\,; \{p_{\ell}\}) - D(p\,; \{p_{\ell}\})$,
\be
t^{(c)}(p\,; \{p_{\ell}\})\, | \lambda_{1}, \ldots, \lambda_{M}\rangle = 
\Lambda^{(c)}(p\,; \{p_{\ell}\},\{\lambda_{j}\})\,
| \lambda_{1}, \ldots, \lambda_{M}\rangle \,,
\ee
with eigenvalue
\be
\Lambda^{(c)}(p\,; \{p_{\ell}\},\{\lambda_{j}\}) = \Lambda_{0}^{(c)}(p\,; 
\{p_{\ell}\})\,
A^{(c)}(p\,; \{\lambda_{j}\}) \,, 
\label{LambdaClosed}
\ee
where $\Lambda_{0}^{(c)}(p\,; \{p_{\ell}\})$ is the pseudovacuum eigenvalue
\be
\Lambda_{0}^{(c)}(p\,; \{p_{\ell}\}) = 
\prod_{\ell=1}^{L}\left( x^{+}(p)-x^{-}(p_{\ell}) \right)
- \prod_{\ell=1}^{L}\left( x^{+}(p)-x^{+}(p_{\ell}) \right)  \,,
\label{Lambda0Closed}
\ee
$A^{(c)}(p\,; \{\lambda_{j}\})$ is the ``dressing'' factor
\be
A^{(c)}(p\,; \{\lambda_{j}\}) = \prod_{j=1}^{M} \left( \frac{x^{-}(p)-x^{+}(\lambda_{j})}
{x^{+}(p)-x^{+}(\lambda_{j})} \right) \,,
\label{dressingClosed}
\ee
and $\{\lambda_{j}\}$ are solutions of the closed-chain BAEs
\be
\prod_{\ell=1}^{L}\left( \frac{x^{+}(\lambda_{j})-x^{-}(p_{\ell})}
{x^{+}(\lambda_{j})-x^{+}(p_{\ell})} \right) = 1\,, \qquad
j = 1, \ldots, M\,, \quad M = 0, 1, \ldots, L-1\,.
\label{BAEClosed}
\ee
We have checked the completeness of this solution numerically for up
to $L=4$.

In the weak-coupling limit (\ref{WeakCoupling}) with also
$x^{\pm}_{\ell} = \frac{1}{g}(u_{\ell} \pm i/2)$, the closed-chain transfer matrix 
becomes \footnote{We suppress the overall factor $1/g^{L}$.}
\be
t^{(c)}(u\,; \{u_{\ell}\}) =  \str_{0} S_{0 L}(u - u_{L}) \cdots S_{0 
1}(u - u_{1}) \,,
\ee
where the $S$-matrix is now given by (\ref{bulkSweak}); and the solution
(\ref{LambdaClosed})-(\ref{BAEClosed}) becomes
\be
\Lambda(u\,; \{u_{\ell}\}, \{\mu_{j}\}) &=&
\left[\prod_{\ell=1}^{L}\left( u - u_{\ell} + i \right)
- \prod_{\ell=1}^{L}\left(u - u_{\ell} \right) \right] 
\prod_{j=1}^{M} \left( \frac{u -\mu_{j} -i}{u -\mu_{j}} \right)\,,  
\non \\
&& \prod_{\ell=1}^{L}\left( \frac{\mu_{j}-u_{\ell}+i}
 {\mu_{j}-u_{\ell}} \right) = 1\,,
\ee
where we have also set $\lambda_{j} =\mu_{j}/g$. This weak coupling 
limit reproduces the results obtained for periodic spin chains based 
on $SU(1|1)$ super-Yangian, see e.g. \cite{Kulish,Kaz,RS,BR}.

We note that the closed-chain transfer matrix (\ref{transferClosed})
does {\it not} satisfy an inversion identity of the form 
(\ref{inversionIntro}). Nevertheless, it does satisfy the relation
\be
t^{(c)}(p\,; \{p_{\ell}\})\, \tilde t^{(c)}(p\,; \{p_{\ell}\}) = 
\Lambda_{0}^{(c)}(p\,; \{p_{\ell}\})\, \tilde \Lambda_{0}^{(c)}(p\,; 
\{p_{\ell}\})\, \id \,,
\ee
where the tilde $(\ \tilde{}\ )$ means that one should make the 
replacement $x^{\pm}(p) \rightarrow x^{\mp}(p)$. The corresponding 
relation for the eigenvalues is evidently satisfied by the expression 
(\ref{LambdaClosed}).

\pagebreak[3]

\section{Pseudovacuum eigenvalue\label{app:pseudovacuum}}

We argue here that the pseudovacuum state $\Omega$ (\ref{pseudovacuum})
is an eigenstate of the open-chain transfer matrix $t(p\,; 
\{p_{\ell}\})$ (\ref{transfer}),
and we compute the corresponding eigenvalue $\Lambda_{0}(p\,; 
\{p_{\ell}\})$.

A direct calculation, usual in the context of open spin chain models, shows that 
$\widehat T_{0}(p\,; \{p_{\ell}\})\, \Omega$ is an upper 
triangular matrix. Then, a careful calculation of  
$R_{0}^{+}(p)\, {\cal T}^{-}_{0}(p\,; 
\{p_{\ell}\}) \, \Omega$ shows, after taking the supertrace 
in the auxiliary space 0, that $\Omega$ is an eigenvector
of the transfer matrix with eigenvalue
\begin{eqnarray}
\Lambda_{0}(p\,; 
\{p_{\ell}\}) &=& (a-x^+)(b+x^-)\prod_{\ell=1}^L 
(x^+-x^-_{\ell})(x^++x^+_{\ell})\nonu
&& - (a+x^-)(b-x^+)\prod_{\ell=1}^L 
(x^+-x^+_{\ell})(x^++x^-_{\ell}) \non \\
&&-\sum_{\ell=1}^L \Bigg\{
(a-x^+)(b-x^+)(x^+-x^-)(x^+_{\ell}-x^-_{\ell})
\nonu
&&\qquad\times
\prod_{k=1}^{\ell-1} (x^+-x^+_{k})(x^++x^-_{k})
\prod_{k=\ell+1}^{L} (x^+-x^-_{k})(x^++x^+_{k})\Bigg\} \,.
\label{eq:L0}
\end{eqnarray}
We have used the notations
\begin{equation}
x^\pm=x^\pm(p) \mb{and} x^\pm_{\ell}=x^\pm(p_{\ell})\,,\quad  
\ell=1,\ldots,L\,.
\end{equation}
This expression can be simplified in the following way. One first 
shows by recursion on $L$ that for any set of variables $y$ and $z^\pm_{\ell}$,
$\ell=1,\ldots,L$, one has the identity
\begin{eqnarray}
2y\,\sum_{\ell=1}^L (z^+_{\ell}-z^-_{\ell})
\prod_{k=1}^{\ell-1} (y-z^+_{k})(y+z^-_{k})
\prod_{k=\ell+1}^{L} (y-z^-_{k})(y+z^+_{k})
\nonu
\ =\ 
\prod_{\ell=1}^L(y-z^-_{\ell})(y+z^+_{\ell})
-\prod_{\ell=1}^L(y-z^+_{\ell})(y+z^-_{\ell}) \,.
\end{eqnarray}
Then, using this relation, one rewrites (\ref{eq:L0}) as
\begin{eqnarray}
\Lambda_{0}(p\,; 
\{p_{\ell}\}) &=& \frac{x^++x^-}{2x^+}\,\left\{
(a-x^+)(b+x^+)\prod_{\ell=1}^L (x^+-x^-_{\ell})(x^++x^+_{\ell})
\right.\nonu
&&\qquad\qquad
 \left.-(a+x^+)(b-x^+)\prod_{\ell=1}^L (x^+-x^+_{\ell})(x^++x^-_{\ell})
\right\} \,.
\end{eqnarray}

\section{Inversion identity}\label{app:inversion}

We write the monodromy matrix with auxiliary space 0 and quantum 
spaces $1,2,\ldots,L$ as a matrix in space 0:
\begin{eqnarray}
\cT_{0}^{(L)}(p)&\equiv& \cT_{0}(p;\{p_{\ell}\}) = 
S_{0L}(p,p_{L})\cdots S_{01}(p,p_{1})\,R_{0}^-(p)\,S_{10}(p_{1},-p)\cdots
S_{L0}(p_{L},-p)\nonu
&=& \left(\begin{array}{cc} A_{1\ldots L}(p) & B_{1\ldots L}(p) 
\\ C_{1\ldots L}(p) & D_{1\ldots L}(p) 
 \end{array}\right) \,.
\end{eqnarray}
We remark that the monodromy matrix is unitary,
\begin{eqnarray}
&& \cT_{0}^{(L)}(p)\,\cT_{0}^{(L)}(-p) = \rho_{L}(p)\,\II_{01\ldots 
L}\,, \label{eq:unit-mono}\\
&& \rho_{L}(p) = (a-x^+)(a+x^-)\,\prod_{\ell=1}^L (x^++x^+_{\ell})
(x^+-x^-_{\ell})(x^-+x^-_{\ell})(x^--x^+_{\ell}) \,,
\end{eqnarray}
which in components reads
\begin{eqnarray}
&& A_{1\ldots L}(p)\,A_{1\ldots L}(-p)+B_{1\ldots L}(p)\,C_{1\ldots L}(-p)
= \rho_{L}(p)\,\II_{1\ldots L}\,,
\label{eq:unitAA}\\
&& D_{1\ldots L}(p)\,D_{1\ldots L}(-p)+C_{1\ldots L}(p)\,B_{1\ldots L}(-p)
= \rho_{L}(p)\,\II_{1\ldots L}\,, \\
&& A_{1\ldots L}(p)\,B_{1\ldots L}(-p)+B_{1\ldots L}(p)\,D_{1\ldots L}(-p)
= 0\,, \\
&& C_{1\ldots L}(p)\,A_{1\ldots L}(-p)+D_{1\ldots L}(p)\,C_{1\ldots L}(-p)
= 0 \,.
\label{eq:unitCA}
\end{eqnarray}
We decompose the scattering matrix in the same way:
\begin{eqnarray}
S_{0L}(p,p_{L}) = \left(\begin{array}{cc} a_{L}(p,p_{L}) & b_{L}(p,p_{L}) 
\\ c_{L}(p,p_{L}) & d_{L}(p,p_{L})  \end{array}\right) 
\equiv
\left(\begin{array}{cc} a_{L}(p) & b_{L}(p) 
\\ c_{L}(p) & d_{L}(p)
 \end{array}\right) \,,
\end{eqnarray}
where we have defined the $2\times 2$ matrices 
\begin{eqnarray}
a_{L}(p,p_{L}) = \left(\begin{array}{cc} \alpha_{1}(p,p_{L}) & 0 \\
0 & \alpha_{2}(p,p_{L}) \end{array}\right)
\mb{;} b_{L}(p,p_{L}) = \left(\begin{array}{cc} 0 & 0\\
\beta(p,p_{L})  & 0\end{array}\right) \label{eq:aL}\\
c_{L}(p,p_{L}) = \left(\begin{array}{cc} 0 & \gamma(p,p_{L}) \\
0 &0 \end{array}\right) \mb{;} 
d_{L}(p,p_{L}) = \left(\begin{array}{cc} \delta_{1}(p,p_{L}) & 0 \\
0 & \delta_{2}(p,p_{L}) \end{array}\right)\label{eq:dL}
\end{eqnarray}
with
\begin{eqnarray}
\alpha_{1}(p,p_{L})= x^+(p)-x^-(p_{L})\ ;\ \alpha_{2}(p,p_{L})= 
x^-(p)-x^-(p_{L}) \label{eq:alph}\\
\beta(p,p_{L})= x^+(p)-x^-(p)\ ;\ \gamma(p,p_{L})= 
x^+(p_{L})-x^-(p_{L})\\
\delta_{1}(p,p_{L})= x^+(p)-x^+(p_{L})\ ;\ \delta_{2}(p,p_{L})= 
x^-(p)-x^+(p_{L})\label{eq:delt}
\end{eqnarray}
or in the weak coupling limit (\ref{WeakCoupling}),
\begin{eqnarray}
&&\alpha_{1}(p,p_{L})= \frac{u-u_{L}+i}{g}\ ;\ \alpha_{2}(p,p_{L})= \frac{u-u_{L}}{g}\ ;\
\beta(p,p_{L})= \frac{i}{g}= \gamma(p,p_{L})\label{eq:alphWk}\\
&&\delta_{1}(p,p_{L})= \frac{u-u_{L}}{g}\ ;\ \delta_{2}(p,p_{L})= 
\frac{u-u_{L}-i}{g}\label{eq:deltWk}
\end{eqnarray}

The identity $S_{L 0}(p_{L},p)=S_{0L}(-p,-p_{L})$ leads to the 
decomposition
\begin{eqnarray}
S_{L0}(p_{L},-p) = \left(\begin{array}{cc} \wh a_{L}(p) & \wh b_{L}(p) 
\\ \wh c_{L}(p) & \wh d_{L}(p) 
\end{array}\right)
\end{eqnarray}
where for any function $f(p)\equiv f(p,p_{L})$, we introduced $\wh f(p)\equiv f(p,-p_{L})$.

The unitary relation for the $S$-matrix (\ref{eq:Sunit}) leads to
\begin{eqnarray}
&& a_{L+1}(p)\,\wh a_{L+1}(-p) +b_{L+1}(p)\,\wh c_{L+1}(-p) 
=-(x^+-x^-_{L+1})(x^--x^+_{L+1})\,\II_{L+1}
\label{eq:aa}\\
&& d_{L+1}(p)\,\wh d_{L+1}(-p) +c_{L+1}(p)\,\wh b_{L+1}(-p) 
=-(x^+-x^-_{L+1})(x^--x^+_{L+1})\,\II_{L+1}
\label{eq:dd}\\
&& a_{L+1}(p)\,\wh b_{L+1}(-p) +b_{L+1}(p)\,\wh d_{L+1}(-p) 
=0\label{eq:ab}\\
&& c_{L+1}(p)\,\wh a_{L+1}(-p) +d_{L+1}(p)\,\wh c_{L+1}(-p) 
=0 \label{eq:ca}
\end{eqnarray}
Note that by changing $p_{L+1}$ to $-p_{L+1}$, one gets a new set of 
relations where `hatted' and `unhatted' functions are exchanged. 
For instance, relation (\ref{eq:aa}) leads to
\begin{eqnarray}
&& \wh a_{L+1}(p)\,a_{L+1}(-p) +\wh b_{L+1}(p)\,c_{L+1}(-p) 
=-(x^-+x^-_{L+1})(x^++x^+_{L+1})\,\II_{L+1}
\end{eqnarray}

Then, the fundamental recursion relation
\begin{equation}
\cT_{0}^{(L+1)}(p) = 
S_{0,L+1}(p,p_{L+1})\,\cT_{0}^{(L)}(p)\,S_{L+1,0}(p_{L+1},-p)
\end{equation}
leads to the following relations
\begin{eqnarray}
A_{1\ldots L+1}(p) &=& a_{L+1}(p)\,\wh a_{L+1}(p)\,A_{1\ldots L}(p)
-a_{L+1}(p)\,\wh c_{L+1}(p)\,B_{1\ldots L}(p)\nonu
&& +b_{L+1}(p)\,\wh a_{L+1}(p)\,C_{1\ldots L}(p)
+b_{L+1}(p)\,\wh c_{L+1}(p)\,D_{1\ldots L}(p) \,,
\label{eq:AL+1}\\
B_{1\ldots L+1}(p) &=& a_{L+1}(p)\,\wh b_{L+1}(p)\,A_{1\ldots L}(p)
+a_{L+1}(p)\,\wh d_{L+1}(p)\,B_{1\ldots L}(p)\nonu
&&
 +b_{L+1}(p)\,\wh d_{L+1}(p)\,D_{1\ldots L}(p) \,,
 \label{eq:BL+1}
 \end{eqnarray}
 \begin{eqnarray}
C_{1\ldots L+1}(p) &=& c_{L+1}(p)\,\wh a_{L+1}(p)\,A_{1\ldots L}(p)
 +d_{L+1}(p)\,\wh a_{L+1}(p)\,C_{1\ldots L}(p) \nonu
&&
+d_{L+1}(p)\,\wh c_{L+1}(p)\,D_{1\ldots L}(p) \,,
\label{eq:CL+1}\\
D_{1\ldots L+1}(p) &=& c_{L+1}(p)\,\wh b_{L+1}(p)\,A_{1\ldots L}(p)
+c_{L+1}(p)\,\wh d_{L+1}(p)\,B_{1\ldots L}(p)\nonu
&& -d_{L+1}(p)\,\wh b_{L+1}(p)\,C_{1\ldots L}(p)
+d_{L+1}(p)\,\wh d_{L+1}(p)\,D_{1\ldots L}(p) \,,
\label{eq:DL+1}
\end{eqnarray}
where we have used the property (\ref{eq:bc-nilpot}) below.
Let us stress that, in the r.h.s. of the above expressions, $A$, $B$, 
$C$ and $D$ act in spaces $1,\ldots,L$ while $a$, $b$, $c$, $d$ act 
in space $L+1$.

Using these expressions, it is a long but simple exercise to show by 
recursion on $L$ that one has the following relations:
\begin{eqnarray}
&&A_{1\ldots L}(p)\,D_{1\ldots L}(-p) \sim \II_{1\ldots L}\mb{;}
D_{1\ldots L}(p)\,A_{1\ldots L}(-p) \sim \II_{1\ldots L}
\label{eq:AD}\\
&& A_{1\ldots L}(p)\,A_{1\ldots L}(-p) +
D_{1\ldots L}(p)\,D_{1\ldots L}(-p) \sim \II_{1\ldots L}\\
&&
A_{1\ldots L}(p)\,C_{1\ldots L}(-p)+C_{1\ldots L}(p)\,D_{1\ldots L}(-p)
= 0 \\
&&B_{1\ldots L}(p)\,A_{1\ldots L}(-p)+D_{1\ldots L}(p)\,B_{1\ldots L}(-p)
= 0\label{eq:BA}
\end{eqnarray}
where the symbol $\sim$ denotes equality up to a multiplication by a 
(scalar) function. The case $L=1$ can be checked directly.
We show explicitly the recursion for the first equality, the other 
ones being proven in the same way. 

We suppose that (\ref{eq:AD})-(\ref{eq:BA}) are valid at a given $L$, 
and expand $A_{1\ldots L+1}(p)\,D_{1\ldots L+1}(-p)$ using expressions
(\ref{eq:AL+1})-(\ref{eq:DL+1}). From unitarity relations 
(\ref{eq:unitAA})-(\ref{eq:unitCA}) and recursion 
hypothesis (\ref{eq:AD})-(\ref{eq:BA}), one can eliminate terms $D_{L}(p)B_{L}(-p)$, 
$D_{L}(p)C_{L}(-p)$, 
$B_{L}(p)D_{L}(-p)$, $C_{L}(p)D_{L}(-p)$, $B_{L}(p)C_{L}(-p)$ and $C_{L}(p)B_{L}(-p)$ from any 
expression. We can also use the property
\begin{eqnarray}
&&c(p,p_{1})\,U\,c(q,q_{1})=0=b(p,p_{1})\,U\,b(q,q_{1})\,,\ \forall\ 
p,q,p_{1},q_{1}\mb{for any diagonal 
matrix $U$,} \non\\
\label{eq:bc-nilpot}
\end{eqnarray}
which is a generalization of the nilpotency for the matrices 
$b$ and $c$. In the same way, since $b$ (respectively $c$) are lower (respectively, 
upper) triangular matrices, we have 
\begin{eqnarray}
V\,b(p,p_{1})\,U\,c(q,q_{1})&=& b(p,p_{1})\,U\,c(q,q_{1})\,V
\label{eq:bc-diago}\\
V\,c(p,p_{1})\,U\,b(q,q_{1})&=& c(p,p_{1})\,U\,b(q,q_{1})\,V
\nonu
\forall\ p,q,p_{1},q_{1}&&
\mb{and for any diagonal matrices $U$ and $V$.}
\nonumber
\end{eqnarray}
As a result, one gets
\begin{eqnarray}
&& A_{1\ldots L+1}(p)\,D_{1\ldots L+1}(-p) \ =\  
a_{L+1}(p)\,\wh a_{L+1}(p)\,d_{L+1}(-p)\,\wh d_{L+1}(-p)\,A_{1\ldots 
L}(p)\,D_{1\ldots L}(-p)\nonu
&& -\Big(a_{L+1}(p)\,\wh c_{L+1}(p)\,d_{L+1}(-p)\,\wh b_{L+1}(-p)
+b_{L+1}(p)\,\wh a_{L+1}(p)\,c_{L+1}(-p)\,\wh 
d_{L+1}(-p)\Big)\,\rho_{L}(p)\,\II_{1\ldots L}
\nonu
&& +b_{L+1}(p)\,\Big(\wh a_{L+1}(p)\,c_{L+1}(-p)
+\wh c_{L+1}(p)\,d_{L+1}(-p)\Big)\,\wh d_{L+1}(-p)\,D_{1\ldots 
L}(p)\,D_{1\ldots L}(-p)\nonu
&& +a_{L+1}(p)\,\Big(\wh a_{L+1}(p)\,c_{L+1}(-p)
+\wh c_{L+1}(p)\,d_{L+1}(-p)\Big)\,\wh b_{L+1}(-p)\,A_{1\ldots 
L}(p)\,A_{1\ldots L}(-p)\\
&& +a_{L+1}(p)\,\Big(\wh a_{L+1}(p)\,c_{L+1}(-p)
+\wh c_{L+1}(p)\,d_{L+1}(-p)\Big)\,\wh d_{L+1}(-p)\,A_{1\ldots 
L}(p)\,B_{1\ldots L}(-p)\nonu
&& +b_{L+1}(p)\,\Big(\wh a_{L+1}(p)\,c_{L+1}(-p)
+\wh c_{L+1}(p)\,d_{L+1}(-p)\Big)\,\wh b_{L+1}(-p)\,C_{1\ldots 
L}(p)\,A_{1\ldots L}(-p)\nonu
&&\!\! -\Big(b_{L+1}(p)\,\wh a_{L+1}(p)\,d_{L+1}(-p)\,\wh d_{L+1}(-p)
+a_{L+1}(p)\,\wh a_{L+1}(p)\,d_{L+1}(-p)\,\wh b_{L+1}(-p)\Big)\,A_{1\ldots 
L}(p)\,C_{1\ldots L}(-p) \,. \nonumber
\end{eqnarray}
Using the explicit forms (\ref{eq:alph})-(\ref{eq:delt}) (or
(\ref{eq:alphWk})-(\ref{eq:deltWk}) if one is interested just in the
weak-coupling limit), one can check that the following relations hold:
\begin{eqnarray}
&& \wh a_{L+1}(p)\,c_{L+1}(-p) +\wh c_{L+1}(p)\,d_{L+1}(-p)=0
\\
&&b_{L+1}(p)\,\wh a_{L+1}(p)\,d_{L+1}(-p)\,\wh d_{L+1}(-p)
+ a_{L+1}(p)\,\wh a_{L+1}(p)\,d_{L+1}(-p)\,\wh b_{L+1}(-p)=0
\label{eq:zero}\\
&&a_{L+1}(p)\,\wh c_{L+1}(p)\,d_{L+1}(-p)\,\wh b_{L+1}(-p)
+b_{L+1}(p)\,\wh a_{L+1}(p)\,c_{L+1}(-p)\,\wh 
d_{L+1}(-p)\sim \II_{L+1}\qquad
\label{eq:idL1}\\
&&a_{L+1}(p)\,\wh a_{L+1}(p)\,d_{L+1}(-p)\wh 
d_{L+1}(-p)\sim \II_{L+1}\label{eq:idL2}
\end{eqnarray}
where, as above, the symbol $\sim$ denotes equality up to a multiplication by a 
(scalar) function. This proves that we have
\begin{eqnarray}
&& A_{1\ldots L+1}(p)\,D_{1\ldots L+1}(-p) \ \sim\  \II_{1\ldots L+1} 
\,.
\label{eq:ADrec}
\end{eqnarray}
Let us stress that in proving relation (\ref{eq:ADrec}), the 
identity $\II_{L+1}$ appearing in (\ref{eq:idL1}) and (\ref{eq:idL2}) is essential to 
pass from $\II_{1\ldots L}$ to 
$\II_{1\ldots L+1}$ in the recursion.
Note also that, apart from the relation (\ref{negation}),
the explicit form of $x^\pm(p)$ is not needed in this calculation.

The other relations are proven along the same lines, and reduce to a 
long list of relations on $a$, $b$, $c$ and $d$ that have to be 
fulfilled. We checked all of them. Most of these relations 
are ensured by the unitarity relations (\ref{eq:aa})-(\ref{eq:ca}) and the 
following quadratic ones:
\begin{eqnarray}
&& a_{L+1}(p)\,\wh a_{L+1}(-p) -c_{L+1}(p)\,\wh 
b_{L+1}(-p)=(x^+-x^+_{L+1})(x^-_{L+1}-x^-)\,\II_{L+1}\\
&& d_{L+1}(p)\,\wh d_{L+1}(-p) -b_{L+1}(p)\,\wh 
c_{L+1}(-p)=(x^+-x^+_{L+1})(x^-_{L+1}-x^-)\,\II_{L+1}\\
&& \wh d_{L+1}(p)\,b_{L+1}(-p) +\wh b_{L+1}(p)\,a_{L+1}(-p)=0\\
%&& \wh a_{L+1}(p)\,c_{L+1}(-p) +\wh c_{L+1}(p)\,d_{L+1}(-p)=0\\
&& \wh a_{L+1}(p)\,d_{L+1}(-p) = 
-(x^++x^+_{L+1})(x^-+x^+_{L+1})\,\II_{L+1}
\label{eq:ad}
\end{eqnarray}
For instance, once (\ref{eq:ad}) 
is proved, (\ref{eq:zero}) reduces to the unitarity relation 
(\ref{eq:ab}), and (\ref{eq:idL2}) is trivially satisfied. In reducing 
the number of equations to be satisfied, properties 
(\ref{eq:bc-nilpot}) and (\ref{eq:bc-diago}) need also to be used.

Some quartic relations, similar to (\ref{eq:idL1}), remain to be checked 
directly. We verified all of them, they take two generic forms. To 
describe these two forms, we introduce the notation 
\begin{eqnarray}
g_{}(p) &=& b_{L+1}(p)\mbox{ or } c_{L+1}(p) \\
\big\{h_{}(p),\ell_{}(p)\big\}&=& 
\big\{a_{L+1}(p),d_{L+1}(p)\big\}\mbox{ or } 
\big\{d_{L+1}(p),a_{L+1}(p)\big\}
\end{eqnarray}
Then, one can check that
\begin{eqnarray}
&&h_{}(p)\,\wh g_{}(p)\,a_{}(-p)\,\wh h_{}(-p)
+\ell_{}(p)\,\wh d_{}(p)\,g_{}(-p)\,\wh \ell_{}(-p)
=0\\
&&h_{}(p)\,\wh g_{}(p)\,d_{}(-p)\,\wh h_{}(-p)
+\ell_{}(p)\,\wh a_{}(p)\,g_{}(-p)\,\wh \ell_{}(-p)
=0
\end{eqnarray}
together with
\begin{eqnarray}
&& h_{}(p)\,\wh h_{}(p)\,b_{}(-p)\,\wh c_{}(-p)
+c_{}(p)\,\wh b_{}(p)\,\ell_{}(-p)\,\wh \ell_{}(-p)
\sim \II_{}
\\
&&b_{}(p)\,\wh h_{}(p)\,h_{}(-p)\,\wh c_{}(-p)
+c_{}(p)\,\wh \ell_{}(p)\,\ell_{}(-p)\,\wh b_{}(-p)\sim \II_{}
\\
&&h_{}(p)\,\wh b_{}(p)\,c_{}(-p)\,\wh h_{}(-p)
+\ell_{}(p)\,\wh c_{}(p)\,b_{}(-p)\,\wh \ell_{}(-p)
\sim \II_{}
\end{eqnarray}
These relations also ensure that the relations (\ref{eq:AD})-(\ref{eq:BA})
 are fulfilled for $L=1$.

Once (\ref{eq:AD})-(\ref{eq:BA}) are proven, it is easy to show 
that the transfer matrix
\begin{equation}
t(p;\{p_{\ell}\}) = \big(b+x^-(p)\big)\,A_{1\ldots L}(p) - \big(b-x^+(p)\big)\,D_{1\ldots L}(p)
\end{equation}
obeys the following relation: 
$$t(p;\{p_{\ell}\})\,t(-p;\{p_{\ell}\})\sim \II_{12\ldots L}.$$
 To determine the 
normalization coefficient, we apply this relation 
onto the pseudovacuum $\Omega$. This leads to
\begin{equation}
t(p;\{p_{\ell}\})\,t(-p;\{p_{\ell}\}) = 
\Lambda_{0}(p;\{p_{\ell}\})\,\Lambda_{0}(-p;\{p_{\ell}\})\,\II_{12\ldots L}\,.
\end{equation}

Finally, let us remark that this proof is also valid in the weak
coupling limit, i.e. for an open spin chain based on the $SU(1|1)$
super-Yangian.  Numerical investigations suggest that the inversion
identity may also be valid for open spin chains based on $SU(n|n)$
super-Yangian, with trivial boundary matrices $R^\pm(p)\sim\II$.

\section{Crossing-like relation\label{app:cross}}

We show here that the open-chain transfer matrix obeys the 
crossing-like relation (\ref{crossinglikereltn}). To this end,
we use the $S$-matrix property (\ref{eq:Scross12}) 
to deduce\footnote{To streamline the notation, we omit
the dependence on the inhomogeneity parameters.}
\begin{eqnarray}
T_{0}(p)^{t_{0}t_{1}\ldots t_{L}} &=& 
(-1)^L\,\sigma^y_{L}\ldots\sigma^y_{1}
\sigma^y_{0}\,\wh T_{0}(-p)\,\sigma^y_{0}
\sigma^y_{1}\ldots\sigma^y_{L}\\
\wh T_{0}(p)^{t_{0}t_{1}\ldots t_{L}} &=& (-1)^L\,\sigma^y_{L}\ldots\sigma^y_{1}
\sigma^y_{0}\,T_{0}(-p)\,\sigma^y_{0}
\sigma^y_{1}\ldots\sigma^y_{L}
\end{eqnarray}
This implies that we have
\begin{equation}
t(p)^{t_{1}\ldots t_{L}} = \sigma^y_{L}\ldots\sigma^y_{1}\,
 str_{0}\Big( R^+_{0}(p)\,\sigma^y_{0}\,\wh T_{0}(-p)\,\sigma^y_{0}\,
 R^-_{0}(p)\,\sigma^y_{0}\,T_{0}(-p)\,\sigma^y_{0}\Big)\sigma^y_{1}\ldots\sigma^y_{L}
\end{equation}
Using cyclicity of the supertrace and the property
\begin{equation}
\sigma^y_{0}\, R^\pm_{0}(p)\,\sigma^y_{0}= R^\pm_{0}(-p)
\end{equation}
we get a first relation on the transfer matrix
\begin{equation}
t(p)^{t_{1}\ldots t_{L}} = \sigma^y_{L}\ldots\sigma^y_{1}\,
 t(-p)\,\sigma^y_{1}\ldots\sigma^y_{L}\,.
\label{eq:transp-transf}
\end{equation}

On the other hand, from the relation (\ref{eq:Scross2}), one 
has
\begin{eqnarray}
T_{0}(p)^{t_{1}\ldots t_{L}} &=& \sigma^y_{1}\ldots\sigma^y_{L}\,
 T_{0}(p)\,\Big|_{x^+\leftrightarrow x^-}\,
\sigma^y_{L}\ldots\sigma^y_{1} 
\\
\wh T_{0}(p)^{t_{1}\ldots t_{L}} &=& \sigma^y_{L}\ldots\sigma^y_{1}\,
\wh T_{0}(p)\,\Big|_{x^+\leftrightarrow x^-}\,
\sigma^y_{1}\ldots\sigma^y_{L} 
\end{eqnarray}
so that one can compute
\begin{eqnarray}
t(p)^{t_{1}\ldots t_{L}} &=& 
 str_{0}\Big( \wh T_{0}(p)^{t_{0}t_{1}\ldots t_{L}}\,R^-_{0}(p)\,
 T_{0}(p)^{t_{0}t_{1}\ldots t_{L}}\,R^+_{0}(p)\Big)\nonu
 &=& 
 str_{0}\Big\{ \Big(\wh T_{0}(p)^{t_{0}t_{1}\ldots t_{L}}\,R^-_{0}(p)\Big)^{t_{0}}\,
 \Big(T_{0}(p)^{t_{0}t_{1}\ldots t_{L}}\,R^+_{0}(p)\Big)^{t_{0}}\Big\}\nonu
 &=& 
 str_{0}\Big( R^-_{0}(p)\,\wh T_{0}(p)^{t_{1}\ldots t_{L}}\,R^+_{0}(p)\,
 T_{0}(p)^{t_{1}\ldots t_{L}}\Big)\\
 &=& \sigma^y_{L}\ldots\sigma^y_{1}\
 str_{0}\Big\{ R^-_{0}(p)\,\Big(\wh T_{0}(p)\Big)_{x^+\leftrightarrow x^-}\,R^+_{0}(p)\,
 \Big(T_{0}(p)\Big)_{x^+\leftrightarrow x^-}\Big\}\
 \sigma^y_{1}\ldots\sigma^y_{L} \nonu
 &=& \sigma^y_{L}\ldots\sigma^y_{1}\ 
 str_{0}\Big\{ \Big(\wh T_{0}(p)^{t_{0}}\Big)_{x^+\leftrightarrow x^-}\,R^-_{0}(p)\,
 \Big(T_{0}(p)^{t_{0}}\Big)_{x^+\leftrightarrow 
 x^-}\,R^+_{0}(p)\Big\}\ 
 \sigma^y_{1}\ldots\sigma^y_{L} \nonu
 &=& \sigma^y_{L}\ldots\sigma^y_{1}\ 
 str_{0}\Big\{ \sigma^y_{0}\,\Big(T_{0}(p)
 \Big)_{\atopn{x^+\leftrightarrow x^-}{x_{\ell}^+\leftrightarrow x_{\ell}^-}}
\,\sigma^y_{0}\,R^-_{0}(p)\,
\sigma^y_{0}\,\Big(\wh T_{0}(p)^{t_{0}}
\Big)_{\atopn{x^+\leftrightarrow x^-}{x_{\ell}^+\leftrightarrow x_{\ell}^-}}
\,\sigma^y_{0}\,R^+_{0}(p)\Big\}\ 
 \sigma^y_{1}\ldots\sigma^y_{L}
\nonumber
\end{eqnarray}
Finally, using the relations
\begin{equation}
\sigma^y_{0}\, R^-_{0}(p)\,\sigma^y_{0} = -R^-_{0}(p)
\Big|_{\atopn{x^+\leftrightarrow x^-}{a\rightarrow -a}} 
\mb{and}
\sigma^y_{0}\, R^+_{0}(p)\,\sigma^y_{0} = -R^+_{0}(p)
\Big|_{\atopn{x^+\leftrightarrow x^-}{b\rightarrow -b}} 
\end{equation}
we deduce
\begin{eqnarray}
t(p)^{t_{1}\ldots t_{L}} = \sigma^y_{L}\ldots\sigma^y_{1}\,
t(p)\,\sigma^y_{1}\ldots\sigma^y_{L}\,
\Big|_{\atopn{x^+\leftrightarrow x^-\ ;\ a\rightarrow -a}
{x_{\ell}^\pm\rightarrow -x_{\ell}^\pm\ ;\ b\rightarrow -b}} 
\,.
\end{eqnarray}

Comparing this last equality with the relation 
(\ref{eq:transp-transf}), we arrive at the desired result
\begin{eqnarray}
t(p) = t(p)\,
\Big|_{\atopn{x^\pm\rightarrow -x^\pm\ ;\ a\rightarrow -a}
{x_{\ell}^\pm\rightarrow -x_{\ell}^\pm\ ;\ b\rightarrow -b}} \,.
\label{crossinglikereltn}
\end{eqnarray}
This identity, applied to a transfer matrix eigenvector, leads to the
following relation for the transfer matrix eigenvalue
\begin{eqnarray}
\Lambda(p) = \Lambda(p)\,
\Big|_{\atopn{x^\pm\rightarrow -x^\pm\ ;\ a\rightarrow -a}
{x_{\ell}^\pm\rightarrow -x_{\ell}^\pm\ ;\ b\rightarrow -b}\ ;\ 
x^+(\lambda_{j})\to-x^+(\lambda_{j})} \,.
\label{crossinglikereltn2}
\end{eqnarray}
Note the change of sign of the Bethe roots $x^+(\lambda_{j})$ 
induced by the BAE (\ref{BAE}). Indeed, it is easy to check that if 
$\{x^+(\lambda_{j})\}$ 
is a set of solutions of these BAEs for the parameters $\{x^\pm, x_{\ell}^\pm, 
a, b\}$, then  
the BAE solutions for the parameters $\{-x^\pm, -x_{\ell}^\pm, 
-a, -b\}$ are given by $\{-x^+(\lambda_{j})\}$.

The same calculation can be done for the closed chain. One 
obtains
\begin{eqnarray}
t(p) &=& (-1)^L\,t(p)\,
\Big|_{\atopn{x^\pm\rightarrow -x^\pm}
{x_{\ell}^\pm\rightarrow -x_{\ell}^\pm}} 
\mb{and}
\Lambda(p) \ =\ (-1)^L\,\Lambda(p)\,
\Big|_{\atopn{x^\pm\rightarrow -x^\pm\ ;\ 
x_{\ell}^\pm\rightarrow -x_{\ell}^\pm}
{x^+(\lambda_{j})\to-x^+(\lambda_{j})}} 
\,.
\end{eqnarray}

\end{appendix}

\end{document}